\input epsf
\input harvmac  
%\writetoc\writedefs
%\noblackbox
%\draftmode\baselineskip=16pt plus 2pt minus 1pt

\Title{\vbox{\hbox{ Brown-HET-1187}}}
{\vbox{
\centerline{  Discrete Spectrum of the Graviton    }
\centerline{ in the $AdS^5$ Black Hole
Background  }  }}

\centerline{Richard C. Brower}
\smallskip
\centerline{Department of Physics, Boston University}
\centerline{Boston, MA 02215, USA}
\bigskip
\centerline{Samir D. Mathur}
\smallskip
\centerline{Center for Theoretical Physics, MIT}
\centerline{Cambridge, MA 02139, USA}
\bigskip
\centerline{Chung-I Tan}
\centerline{Department of Physics, Brown University}
\centerline{Providence, RI 02912, USA}

\vskip .5in

\noindent
The discrete spectrum of fluctuations of the  metric
about an $AdS^5$ black hole background are found. These modes are the
strong coupling limit of so called glueball states in a dual 3-d
Yang-Mills theory with quantum numbers $J^{PC} = 2^{++}, 1^{-+},
0^{++}$. For the ground state modes, we find the mass relation:
$m(0^{++}) < m(2^{++}) < m(1^{-+})$.  Contrary to expectation, the
mass  of our new $0^{++}$ state ($m^2=5.4573$) associated with the 
graviton is smaller than the mass of the $0^{++}$  state ($m^2=11.588$)
from the dilaton.  In fact the dilatonic excitations are exactly
degenerate with our  tensor $2^{++}$  states. We find that 
variational methods gives remarkably accurate mass estimates
for all three low-lying levels while a WKB treatment
describes the higher modes well.

\Date{August 1999}

\vfill\eject

\def\lapl{\,\raise.5pt\hbox{$\mbox{.09}{.09}$}\,}

\def\IR{\relax{\rm I\kern-.18em R}}
\font\cmss=cmss10 \font\cmsss=cmss10 at 7pt
\def\IZ{\relax\ifmmode\mathchoice
{\hbox{\cmss Z\kern-.4em Z}}{\hbox{\cmss Z\kern-.4em Z}}
{\lower.9pt\hbox{\cmsss Z\kern-.4em Z}}
{\lower1.2pt\hbox{\cmsss Z\kern-.4em Z}}\else{\cmss Z\kern-.4em Z}\fi}
\def\dddots{\mathinner{\mkern2mu\raise1pt\vbox{\kern7pt\hbox{.}}\mkern2mu
   \raise4pt\hbox{.}\mkern2mu\raise7pt\hbox{.}\mkern1mu}}

\ifx\epsfbox\UnDeFiNeD\message{(NO epsf.tex, FIGURES WILL BE IGNORED)}
\def\figin#1{\vskip2in}% blank space instead
\else\message{(FIGURES WILL BE INCLUDED)}\def\figin#1{#1}\fi

\def\ifig#1#2#3{\xdef#1{fig.~\the\figno}
\writedef{#1\leftbracket fig.\noexpand~\the\figno}%
\goodbreak\midinsert\figin{\centerline{#3}}\centerline{\vbox{\baselineskip12pt
\advance\hsize by -1truein\noindent\footnotefont{\bf Fig.~\the\figno:} #2}}
\bigskip\endinsert\global\advance\figno by1}

\def\figinsert#1#2#3{\topinsert\figin{#2}\centerline{\vbox{\baselineskip12pt
\advance\hsize by -1truein\noindent\footnotefont{\bf Fig.~\xfig#1:} #3}}
\bigskip\endinsert}

%

%

%

%

%

%

%

%

%Put References Here : This is very nice since it
%puts the right reference in the (correct) order of citations.
%STUPID RESTRICTIONS ON NUMBERS: Replace  "0123456789" by "ZOTRFVXSEN" !
%Also can put in tex but it is then much messier to work with.

\lref\maldacena{J. Maldacena, ``The Large $N$ limit of superconformal
field theories and supergravity'', Adv. Theor. Math. Phys. 2:231, 1998, 
hep-th/9711200.}

\lref\wittenO{ E. Witten, ``Anti-de Sitter space and holography'',
Adv. Theor. Math. Phys.2: 253, 1998, hep-th/9805028.}

\lref\gkp{ S.S. Gubser, I.R. Klebanov and A.M. Polyakov, ``Gauge theory
correlators from noncritical string theory'',  Phys. Lett. {\bf B428} (1998) 
105, hep-th/9802109.}

\lref\wittenT{ E. Witten, ``Anti-de Sitter space, Thermal Phase Transition,
and Confinement in Gauge Theories'',  Adv. Theor. Math. Phys.2: 505, 1998, 
hep-th/9803131}

\lref\csaki{ C. Cs\'aki, H. Ooguri, Y. Oz and J. Terning, ``Glueball
Mass Spectrum From Supergravity'', hep-th/9806021.}

\lref\jev{R. De Mello Koch, A. Jevicki, M. Mihailescu and J. Nunes, 
``Evaluation of glueball masses from supergravity,'' hep-th/9806125.}

\lref\several{D. Gross and H. Ooguri, Phys. Rev. {\bf D58} (1988) 106002, hep-th/9805129,  M. Zyskin, 
Phys. Lett. {\bf B439} (1998) 373, hep-th/9806128,  H. Oooguri, H. Robins
and J. Tannenhauser, Phys. Lett. {\bf B437} (1998) 77, hep-th/9806171, J.G. Russo, hep-th/9808117,  A. Hashimoto and Y. Oz, hep-th/9809106, C. Cs\'aki, H. Ooguri, Y. Oz and J. Terning,
hep-th/9810186, S.S. Gubser,
hep-th/9810225, P. Kraus, F. Larsen and S.P. Trivedi, hep-th/9811120, R.G. Cai and
K.S. Soh, hep-th/9812121,  J.G. Russo and K. Sfetsos, hep-th/9901056, C. Cs\'aki, J. Russo, K. Sfetsos and J. Terning, hep-th/9902067.}

\lref\tanO{For a recent review, see:  C-I Tan, ``Diffractive
Production at Collider Energies and Factorization", Phys. 
Report, {\bf 315} (1999) 175,  hep-ph/9706276, hep-ph/9810237.}
\lref\tanT{For a review on experimental support for identifying the Pomeron
Regge  trajectory as a
closed-string exchange in large-N QCD, see: A. Capella, U. Sukhatme, C-I Tan,
and J.T.V. Tran, 
Phys. Report, {\bf 236} (1994) 225.}
\lref\tanR{For recent attempts to extra Pomeron in $d=3$ {QCD}, see:  
M. Li and C-I Tan, 
Phys. Rev. {\bf
D51} (1995) 3287; D. Yu. Ivanov, R. Kisrchner, E.M. Levin, L.N. Lipatov, L.
Szymanowski, and M. Wusthoff, Phys. Rev. {\bf D58} (1998) 74010. }
\lref\chandra{S. Chandrasekhar, `The Mathematical theory of Black Holes' (1983)
{\it Oxford University Press}.}

 \lref\rw{T. Regge
and J. A. Wheeler, Phys. Rev. {\bf 108} (1957) 1063.}

\lref\langer{K. E. Langer, Phys. Rev. {\bf 51}, 669 (1937).}

\lref\minahan { J. Minahan, ``Glueball mass spectra and other issues for 
supergravity duals of QCD models,'' hep-th/9811156.}

\lref\cm{N. R. Constable and R. C. Myers, hep-th/9908175.}

\newsec{Introduction}

The Maldacena conjecture~\maldacena~ and its further
extensions~\wittenO\gkp~ allow us to compute quantities in a strongly
coupled gauge theory from its dual gravity description. In particular,
Witten~\wittenT~ has pointed out if we compactify the 4-dimensional
conformal super Yang Mills (SYM) to 3 dimensions using anti-periodic
boundary conditions on the fermions, then we  break supersymmetry and
conformal invariance and obtain a theory that has interesting mass
scales. In Refs. \csaki\ and \jev, this approach was used to calculate
 a discrete
mass spectrum for $0^{++}$ states associated with $Tr[F^2]$ at strong
coupling by solving the dilaton's wave equation in the corresponding
gravity description. Although the theory at strong coupling is 
really not pure Yang-Mills, since it has additional fields,
some rough agreement was claimed between the pattern of glueball
masses thus obtained and the masses computed for ordinary 3-d
Yang-Mills by lattice calculations.  In Ref.~\csaki~ the calculation
was extended to solving the modes of the vector and two form fields
with certain polarizations. Several other investigations have expanded
our knowledge of the relation between string theory and gauge
theories, in these and related directions \several.

Here we consider the discrete spectrum for the graviton field which is
dual to the energy momentum tensor of the Yang-Mills field theory.
This is a very interesting case, since a 5-d graviton field has
five independent on-shell states that in a dual correspondence to a
4-d Yang-Mills theory would account for the five helicity components
of a massive $2^{++}$ state lying on the leading diffractive (Pomeron)
trajectory~\tanO\tanT. In the present analysis, these five states are 
mapped by
dimensional reduction to a reducible multiplet with quantum numbers
$2^{++}$, $1^{-+}$, $0^{++}$ in a 3-d Yang Mills theory~\tanR. Contrary to
expectations, we find that the mass gap defined by the lowest $0^{++}$
state in the graviton multiplet is lower than mass gap for the
dilaton. In fact our $2^{++}$ states are exactly degenerate with the
lowest dilaton states, which shall be denoted as ${\tilde 0}^{++}$, 
while the $1^{-+}$ equation gives the most
massive states. Remarkably these mass relations, 
$$m(0^{++}) <
 m({\tilde 0}^{++}) = m(2^{++}) < m(1^{-+}),$$ 
for the lowest modes are reminiscent of
weak coupling glueball spectra from lattice gauge theory or the MIT
bag model. Extensions to the graviton multiplet for states dual to the
4-d Yang Mills theory will be reported separately.

Quite apart from the issue of how closely the gauge theory obtained
from the AdS description at strong coupling agrees ``accidentally''
with ordinary Yang-Mills, it is a very interesting fact that one
obtains discrete masses representing long lived states in the theory
from a very different looking calculation done with the gravity
variables. On the gravity side, if we compactify the $AdS_5$ metric on a large
$S^1$ then the metric is a periodic version of the original $AdS^5
\times S^5$ metric. At high temperature (small radius of the $S^1$) this
metric is replaced by  an AdS black hole metric.  This geometry has
stationary modes for the fields of string theory (and in particular
the fields of supergravity), and it is conjectured that the spectrum
of these modes gives the mass spectrum for the compactified gauge
theory on $R^3 \times S^1$.  Understanding the properties of such a
Yang-Mills theory even at strong coupling is interesting.

In this paper we carry out the calculation of the discrete modes for
the perturbations of the gravitational metric. We restrict the
perturbations to be independent of the $S^5$ factor in the metric
and constant in the co-ordinate for the compact $S^1$. Our results are
presented as follows.

In sec II, we derive three different wave equations that arise from
fluctuations in the gravitational metric.  These correspond to spin-2,
spin-1 and spin-0 perturbations from the viewpoint of the rotation
group in the 3-dimensional noncompact space:
 
(i)\quad The spin-2 equation can be mapped by a change of variables to
the equation for a massless scalar, and so the energy levels found are
the same as those for the dilaton field $\phi$.

(ii)\quad The spin-1 equation gives energy levels that are somewhat higher 
than the spin-2 case.

(iii)\quad We first obtain the spin-0 equation in a gauge which has only
diagonal components of the metric. We obtain a third order  equation for the
perturbation. We observe that there is a one parameter residual gauge freedom
left in the ansatz, and we use the form of this gauge mode to reduce the third
order equation to a second order equation for the perturbation.  The situation 
here is similar to the
analysis of polar perturbations of the 3+1 Schwarzschild hole in a
 diagonal gauge, where  a third order
system is also found. We comment on the gauge freedom present in
the latter case; this discussion may illuminate some of the mysteries
associated with the existence of the second order `Zerilli equation'~\chandra\
for the above mentioned polar perturbations of the Schwarzschild
metric. As a check on our result, we then study the perturbation 
 in an alternative gauge,
similar to one used by Regge and Wheeler~\rw. In this latter gauge we directly 
obtain a second order
equation, which is seen to be equivalent to the equation obtained by the
first method.

In Sec III, we solve these equations numerically. We comment of the
issue of the boundary condition to be used at the horizon of the black
hole: The physics we impose is that there be no flux transfer across
the horizon of the black hole, in agreement with the boundary condition
used in Ref.~\jev~. (In Appendix B, we also show that
the desired boundary conditions at $r=1$ and $r=\infty$ together  allow us to
treat  these equations  in a Sturm-Liouville approach.) We present accurate
numerical solutions to the first 10 levels.  The spin-0 equation gives energy
levels that are lower than the spin-1 and spin-2 equations. 
 
In Sec IV, we consider approximation methods to gain insight into the
mass relations. The ground state mass is also given in terms of a very
simple variational ansatz and the entire spectrum approximated very
well by a low-order WKB approximation. We also demonstrate how the
boundary condition at the horizon can be implemented more
effectively with the help of an ``effective angular
momentum". (Technical details are postponed to Appendix B on the
variational method, Appendix C on the Schrodinger form of the wave
equations and Appendix D on the derivation of the WKB expansion.)

In sec V, we discuss the results and make a few cautionary remarks on
the difficulties of the comparison of this strong coupling spectrum
with glueballs as computed in lattice gauge theory or classified in
bag models. We recommend a more thorough analysis of the complete set of
spin-parity states for the entire bosonic supergravity multiplet and its
extension to 4-d Yang Mills models.

\subsec{\bf AdS/CFT correspondence at Finite Temperature}

Let us review briefly the proposal for getting a 3-d Yang-Mills theory
dual to supergravity. One begins by considering Type IIB supergravity
in Euclidean 10-dimensional spacetime with the topology 
$M_5\times S^5$.  The  Maldacena conjecture asserts that
IIB superstring theory on $AdS^5\times S^5$ is dual to the
${\cal N } = 4$ SYM conformal field theory on the boundary of the $AdS$ space.
The metric of this spacetime is
\eqn\ads{ds^2/R^2_{ads} =  r^2 (d\tau^2+ dx^2_1 + dx^2_2 + dx^2_3) +
{ dr^2 \over r^2 } + d\Omega^2_5 \; ,}
where the radius of the $AdS$ spacetime is given through
$R^4_{AdS} = g_s N \alpha'^2$ ($g_s$ is the string coupling and
 $l_s$ is the string length, $l^2_s = \alpha'$). The Euclidean time is 
 $\tau = i
x_0$.  To break conformal invariance, 
 following \wittenT , we
place the system at a nonzero temperature described by a  periodic Euclidean
time $\tau = \tau + 2 \pi R_0$. The metric correspondingly
changes, for small enough $R_0$, to  the non-extremal black hole
metric in $AdS$ space. For large black hole temperatures, the stable phase
of the metric corresponds to a black hole with radius large compared to the
$AdS$ curvature scale. To see the physics of discrete modes we may take the
limit of going close to the horizon of this large hole, whereupon the metric
reduces to that of the black 3-brane. This metric is (we scale out all
dimensionful quantities)
\eqn\three{ds^2=f(r)d\tau^2+f^{-1}(r)dr^2+r^2(dx_1^1+dx_2^2+dx_3^2)+
d\Omega_5^2  \; , } 
where
\eqn\threep{f(r)=r^2-{1\over r^2} }

On the gauge theory side, we have a $N=4$ susy theory corresponding
to the $AdS$ spacetime, but with the $S^1$ compactification we have this
theory on a circle with antiperiodic boundary conditions for the fermions. 
Thus
supersymmetry is broken and massless scalars are expected to acquire
quantum corrections.

 From the view point of a 3-d theory, 
 the compactification radius acts as
an UV cut-off. Before the compactification the 4-d theory was
conformal, and was characterized by a dimensionless effective coupling
 $({g_{YM}^{(4)}})^2N\sim
g_s N$. After the compactification the theory is not conformal, and the radius
of the compact circle provides a length scale. Let this radius be $R$. Then a
naive dimensional reduction from 4-d Yang-Mills to 3-d Yang-Mills,
would give an effective coupling in the 3-d theory equal to
$(g_{YM}^{(3)})^2N=({g_{YM}^{(4)}})^2N/ R$. This 3-d coupling has the
 units of mass. If the dimensionless coupling $({g_{YM}^{(4)}})^2N$ is much
 less than unity, then the length scale associated to this mass is larger than
 the radius of compactification, and we may expect the 3-d theory to
be a
 dimensionally reduced version of the 4-d theory. Unfortunately the
dual
 supergravity description applies at $({g_{YM}^{(4)}})^2N>>1$, so that
 the higher Kaluza-Klein modes of the $S^1$ compactification have lower energy
 than the mass scale set by the 3-d coupling. Thus we do not really
have a 
 3-d gauge theory with a finite number of additional fields.
 
 One may nevertheless expect that some general properties of the dimensionally
 reduced
 theory might survive the change between small and large coupling. For this
 purpose we
 look at the pattern of masses and spins that are obtained for the fields that
 are singlets under the $S^1$ - i.e. we ignore the Kaluza-Klein modes of the
 $S^1$. In keeping with earlier work, we also restrict ourselves to modes 
 that are singlets of the $SO(6)$, since  non-singlets under the $S^1$ and
 the $SO(6)$ can have no counterparts in a dimensionally reduced $QCD_3$.

\newsec{Wave Equations}

\subsec{\bf Field equations and the equilibrium configuration}
We consider Type IIB supergravity in Euclidean 10-dimensional
spacetime. The spacetime will have the topological form $M_5\times
S^5$. For the perturbations that we are interested in, we can set to
zero the dilaton $\phi$, the axion $C$ and the 2-form fields
$B^{NSNS}, B^{RR}$. The metric and the 4-form gauge field satisfy the
Einstein equation
\eqn\zone{R_{\hat  \mu\hat  \nu}={1\over 6}
F_{\hat\mu\hat\rho\hat\sigma\hat\tau\hat\kappa}
F_{\hat\nu}^{\hat\rho\hat\sigma\hat\tau\hat\kappa} }
and the self-duality condition for the field strength of the four form
\eqn\ztwo{F_{\hat\mu\hat\nu\hat\rho\hat\sigma\hat\tau}=
{1\over 5!}
\epsilon_{\hat\mu\hat\nu\hat\rho\hat\sigma\hat\tau
\hat\mu'\hat\nu'\hat\rho'\hat\sigma'\hat\tau'}
F^{\hat\mu'\hat\nu'\hat\rho'\hat\sigma'\hat\tau'}
} {\it\bf Notations}:\quad Indices with a `hat' are 10-d indices. 
The coordinates of the $S^5$ will
be called $y^\alpha$, and the indices here will be $\alpha, \beta, \dots$. 
The coordinates of $M$
will be
$x^\mu$, and the indices will be $\mu, \nu, \dots$. The $\epsilon$
tensor is real in spacetime with Lorentzian signature, and thus pure
imaginary in Euclidean signature. Thus $F$ will be real on $S^5$, and
imaginary on $M_5$. 

The equilibrium configuration about which we perturb will be given by the 
metric
\eqn\three{ds^2=f(r)d\tau^2+f^{-1}(r)dr^2+r^2(dx_1^1+dx_2^2+dx_3^2)+
d\Omega_5^2}
where
\eqn\threep{f(r)=r^2-{1\over r^2}}
The last term in \three\ gives the metric of $S^5$ and we have chosen
units to make this a sphere of unit radius. The 5-form field strength
with indices in the $S^5$ is a constant times the volume form on
$S^5$. By the self duality condition \ztwo\ we have
\eqn\four{F_{  \mu  \rho  \sigma  \tau  \kappa}F_{  \nu}^{  \rho  \sigma  
\tau  \kappa}=-2\Lambda g_{\mu\nu}}
where
\eqn\five{2\Lambda={1\over 5}F_{  \alpha  \beta  \gamma  \delta  \epsilon}
F^{  \alpha  \beta  \gamma  \delta  \epsilon}=-{1\over 5}F_{  \lambda  \rho  
\sigma  \tau  \kappa}F^{  \lambda  \rho  \sigma  \tau  \kappa}}
is a positive constant.

\subsec{\bf Ansatz for the perturbations}

We wish to consider fluctuations of the metric of the form 
\eqn\six{g_{\mu\nu}=\bar g_{\mu\nu}+ h_{\mu\nu}(x)}
Thus the perturbations will have no dependence on the coordinates of
the $S^5$. Further, we wish to keep unchanged all the other fields in
the theory. It is easy to see that $\phi, C, B^{NSNS}, B^{RR}$ can be
held fixed, since they arise quadratically in the action, and vanish
in the equilibrium configuration. (The metric above is the Einstein
metric, and so there is no linear coupling of $h$ to the dilaton.) It
is also consistent to hold fixed the geometry of the $S^5$ and the
5-form field-strength on the $S^5$.  Keeping these quantities fixed
keeps unchanged the flux of $F$ on $S^5$, which is one constraint that
the perturbations must respect. The self duality condition \ztwo\ can
then regarded as determining the value of $F$ on $M$. But then we see
that we will still get after such a perturbation:
\eqn\seven{ F_{  \mu  \rho  \sigma  \tau  \kappa}F_{  \nu}^{  \rho  \sigma  
\tau  \kappa}
=-2\Lambda g_{\mu\nu}}
with $\Lambda$ the same constant as in \five .
The Einstein equations then give
\eqn\zonep{R_{\mu\nu}=-{1\over 3}\Lambda g_{\mu\nu}}
Here the Ricci tensor $R_{\mu\nu}$ can be computed solely from the 5-d
metric of $M$; all contributions involving the $S^5$ variables are
zero. Thus if we find a solution to \zonep , we have a consistent
solution the to entire supergravity problem. With the expansion \six\
we get
\eqn\thir{-{1\over 2}h_{\mu\nu;\lambda}{}^{\lambda}
-{1\over 2}h_\lambda^\lambda{}_{;\mu\nu}
+{1\over 2}h_{\mu\lambda;\nu}{}^{\lambda}
+{1\over 2}h_{\nu\lambda;\mu}{}^{\lambda}+(D-1)h_{\mu\nu}=0}
where $D=5$.

To see how many kinds of modes we expect, consider first the graviton in 5-d 
flat spacetime with coordinates $x^i, i=1,\dots 5$, and signature $(-++++)$. 
Then the graviton modes can be brought by gauge freedom to a form
\eqn\eight{h_{\mu\nu}=\epsilon_{\mu\nu}e^{ik(x_2-x_1)}, ~~~\mu, 
\nu=3,4,5, ~~~\epsilon_\mu^\mu=0}
Thus there are 5 independent polarizations of the graviton.

In the spacetime \three\ the coordinates $x_1, x_2, x_3$ project to
the 3-d  spacetime of the gauge theory that we wish to study, $\tau$
is direction of  the gauge theory that is compactified in obtaining
the 3-d theory from the  4-d theory, and $r$ is the radial coordinate
of the spacetime that loosely  speaking governs the scale at which the
gauge theory is being studied. Our perturbations will have the form
\eqn\ten{h_{\mu\nu}=\epsilon_{\mu\nu}(r)e^{-mx_3}}
where we have chosen to use $x_3$ as a Euclidean time direction to define 
the glueball masses of the 3-d gauge theory. We start with fixing for the 
gravitational perturbations the gauge
\eqn\el{h_{3\mu}=0}
though later we will have occasion to consider changes to another gauge as 
well.

\subsec{\bf Wave equations for the perturbations}

 From the above ansatz and the metric \three\ we see that we have an 
$SO(2)$ 
rotational symmetry in
the $x_1-x_2$ space, and we can classify our perturbations with respect to 
this symmetry. We then
find the following categories of perturbations:
\item{(a)} Spin-2: There are two linearly independent perturbations which 
form the spin-2
representation of the above $SO(2)$:
\eqn\tw{h_{12}=h_{21}=q_T(r)e^{-mx_3}, ~~~{\rm all ~other ~components~zero} }
\eqn\tw{h_{11}=-h_{22}=q_T(r)e^{-mx_3}, ~~~{\rm all ~other~components~ zero} }
The Einstein equations \thir\ give,
\eqn\fourt{r^2(1-{1\over r^4})q_T''(r)+r(1+{3\over r^4})q_T'(r)+
[{m^2\over r^2}-4(1+{1\over
r^4})]q_T(r)=0.} Defining
\eqn\fift{\phi_T={q_T \over r^2} }
Eq. \fourt\ becomes
\eqn\fift{r^2(1-{1\over r^4})\phi_T''(r)+r(5-{1\over r^4})\phi_T'(r)
+{m^2\over r^2}\phi_T(r)=
\phi_{T;\lambda}{}^\lambda=0,}
which is the free wave equation, and thus the same equation as that satisfied 
by 
the dilaton
(with constant value on the $S^5$).
These dilaton modes have been studied in \csaki \jev\ for example , and we therefore will 
obtain the same 
energy
levels for the spin-2 perturbations as those obtained for the dilaton, 
{\it e.g.}, for the ground
state, $m^2_{T,0} \simeq 11.588.$ 
\item{(b)} Spin-1:
\eqn\tw{h_{i\tau}=h_{\tau i}=q_V(r)e^{-mx_3}, {i=1~~or ~~i=2},
 ~~~{\rm all ~other~components~zero} }
The Einstein equations \thir\ give, 
\eqn\spinOne{r^2(1-{1\over r^4})q_V''(r)+r(1-{1\over r^4})q_V'(r)+
[{m^2\over r^2}-4(1-{1\over
r^4})]q_V(r)=0}
\item{(c)} Spin-0: Based on the symmetries we choose an ansatz where the 
nonzero components of the
perturbation are
\eqn\sixt{\eqalign{h_{11}&=h_{22}=q_1(r)e^{-mx_3}\cr
h_{\tau\tau}&=-2q_1(r){f(r)\over r^2}e^{-mx_3}+q_2(r)e^{-mx_3}\cr
h_{rr}&=q_3(r)e^{-mx_3}\cr }}
where $f(r)$ is given in \threep . We have chosen to express $h_{\tau\tau}$ 
in such a  way so that 
the part involving $q_1$ is traceless, which we expect is to dominate at large 
$r$.
\item{}The analysis of the Einstein equations are given in Appendix A, where
 they are derived in the
above gauge and also in a different gauge which is less directly related to
 the flat space form,
\eight, but more easy to solve:
\eqn\sixtq{\eqalign{h_{\tau\tau}&=q_1(r)e^{-mx_3}\cr
h_{rr}&=q_2(r)e^{-mx_3}\cr
h_{r3}=h_{3r}&=q_3(r)e^{-mx_3}\cr }}
The field equation is, for $q_3\equiv q_S(r)$,
\eqn\thtreeq {\big\{p_2(r) {d^2\over dr^2} + p_1(r) {d\over dr} + p_0(r)\big\}
 q_S(r)=0,} 
where
\eqn\ththreeqCoeffs{\eqalign{p_2(r)
&=r^2(r^4-1)^2 [ 3(r^4-1) + m^2 r^2 ]\cr
p_1(r)&=r(r^4-1)[3(r^4-1)(5r^4+3)+m^2r^2(7r^4+5)]\cr
p_0(r)&=9(r^4-1)^3+2m^2r^2(3+2r^4+3r^8)+m^4r^4(r^4-1).\cr}}

\newsec{Numerical Solution}

To calculate the discrete spectrum for our three equation, one must 
apply the correct boundary conditions at $r = 1$ and $r = \infty$.
>From the indicial equations,  the  asymptotic
value of the two linearly independent solutions at infinity are 
\eqn\rone{\eqalign{q_T(r) &\sim  r^{-2}, \; r^2   \cr
q_V(r) &\sim  r^{-2}, \; r^2  , \cr q_S(r) &\sim r^{-3}, \; r^{-1} \;, }}
and at $r= 1$ are
\eqn\rinfty{\eqalign{
q_T(r) &\sim 1, \; log(r-1) \; , \cr q_V(r) &\sim r-1 , \; (r-1)
log(r-1) \; , \cr q_S(r) &\sim (r-1)^{-1}, \; (r-1)^{-1} log(r-1)
\;. }} In all cases the appropriate boundary condition~\jev~ at
$r =1$ is the one without the logarithmic singularity.  At $r =
\infty$ the least singular boundary is required to have a normalizable
eigenstate. Matching these two boundary conditions results in a
discrete set of eigenvalues $m^2_n$, where $n$ is the number of zeros
in the wave function inside the interval $r \in (1,\infty)$.  We
solved the eigenvalue equations  by the shooting method, integrating
from $r_1 \simeq 1$ to large $r_\infty \simeq \infty$.  The parameter $m^2$ is
adjusted by applying Newton's method to satisfy the condition
$\lim_{r\rightarrow \infty} q(r) = 0$.  To begin the integration one needs the function
and its first derivative at $r=1$. Thus expanding to the next
significant order for the regular solution at $r=1$, the initial
conditions for the tensor were taken to be $q_T(r_1) = 1, \; q'_T(r_1) =
2 - m^2/4$ and for the vector to be $q_V(r_1) = 0, \; q'_V(r_1) = 1$.
For the scalar equation, we found it convenient to define an new
function, $f_S(r) = (r^2 -1) q_S(r)/r$, so that the initial value is
nonsingular and only the correct solution at infinity vanishes. This
leads to initial conditions, $f_S(r_1)= 1, \; f'_S(r_1)= -m^2/4$.

Below we present in Table 1 the first 10 states using this shooting
method.  The spin-2 equation is equivalent to the dilaton equation
solved in Refs.~\csaki\ and \jev , so the excellent agreement with earlier values
validates our method.  In all of the equation we use a standard
Mathematica routine with boundaries taken to be $x= r^2 -1 = \epsilon$
and $1/x = \epsilon$ reducing $\epsilon$ gradually to $\epsilon =
10^{-6}$.  Note that since all our eigenfunctions must be even in $r$
with nodes spacing in $x = r^2-1$ of $O(m^2)$, the variable $1/x$ is a
natural way to measure the distance to the boundary at infinity.  For
both boundaries, the values of $\epsilon$ was varied to demonstrate
that they were near enough to $r = 1,
$ and $ \infty$ so as not to
substantially effect the answer.

\midinsert
$$\vbox{\offinterlineskip
        \halign{&\vrule#&\strut\ #\ \cr
\multispan{9}\hfil Glueball $m^2_n$ by Shooting Technique \hfil\cr
\noalign{\medskip}
\noalign{\hrule}
&\hfil\bf level \hfil&&\hfil $J^{PC} = 0^{++}$ \hfil&&\hfil $J^{PC} = 1^{-+}$ 
\hfil&&\hfil $J^{PC} = 2^{++}$ \hfil&
\cr \noalign{\hrule}
&\hfil n= 0  \hfil &&\hfil 5.4573   \hfil&&\hfil 18.676 \hfil&&\hfil 11.588  
\hfil& 
\cr \noalign{\hrule}
&\hfil n= 1  \hfil &&\hfil  30.442  \hfil&&\hfil 47.495 \hfil&&\hfil 34.527 
\hfil& 
\cr \noalign{\hrule}
&\hfil n= 2  \hfil &&\hfil 65.123   \hfil&&\hfil 87.722   \hfil&&\hfil
68.975  \hfil& 
\cr \noalign{\hrule}
&\hfil n= 3  \hfil &&\hfil 111.14     \hfil&&\hfil 139.42  \hfil&&\hfil
114.91 \hfil& 
\cr \noalign{\hrule}
&\hfil n= 4  \hfil &&\hfil 168.60    \hfil&&\hfil 203.99 \hfil&&\hfil
172.33 \hfil& 
\cr \noalign{\hrule}
&\hfil n= 5  \hfil &&\hfil 237.53   \hfil&&\hfil 277.24 \hfil&&\hfil
241.24 \hfil& 
\cr \noalign{\hrule}
&\hfil n= 6  \hfil &&\hfil 317.93   \hfil&&\hfil 363.38 \hfil&&\hfil
321.63 \hfil& 
\cr \noalign{\hrule}
&\hfil n= 7  \hfil &&\hfil  409.82  \hfil&&\hfil 461.00  \hfil&&\hfil
413.50 \hfil& 
\cr \noalign{\hrule}
&\hfil n= 8  \hfil &&\hfil 513.18    \hfil&&\hfil  570.11 \hfil&&\hfil
516.86 \hfil& 
\cr \noalign{\hrule}
&\hfil n= 9  \hfil &&\hfil 628.01   \hfil&&\hfil 690.70 \hfil&&\hfil
631.71 \hfil& 
\cr \noalign{\hrule}
}}$$
\centerline{Table 1: Radial Excitations for Glueballs associated with
the Gravitation tensor field.}
\endinsert
\bigskip

\vskip30pt

\newsec{Main Features of Glueball Spectrum}

There are two intriguing aspects of our numerical results for the glueball 
spectrum. The first 
concerns the pattern of low-mass states, {\it e.g.}, for the respective 
ground states, one has: 
$m^2_{S,0}< m^2_{T,0} < m^2_{V,0}$. The second interesting feature is the 
asymptotic growth for the
masses of their radial excitations. In order to provide further insights 
concerning these features,
we turn next to approximation schemes appropriate for discussing: 
(1)  ground state masses, and (2) 
high mass excitations. We end this section by  providing a heuristic picture 
for understanding
these main features.

 \subsec{\bf Low-Mass Spectrum}

Eq. \fift\ for  tensor modes can be expressed in
 a standard ``Sturm-Liouville" form
\eqn\SL{\Big(-{d\over dx}x(x+1)(x+2){d\over dx} \> +  w(x)
\>\Big ) \phi(x) = {m^2\over 4} \phi(x)\quad,}
where we have found it more convenient to use $x\equiv r^2-1$ as the
new variable. For tensor modes, the potential term is in fact absent,
{\it i.e.}, $w_T(x)=0$.  Eq. \spinOne\ for vector modes can also be
brought into this form, with $\phi_V(x)\equiv q_V(r)/\sqrt
{r^4-1}$ and $w_V(x)= 1/(x (x+1)(x+2))$.  As a Sturm-Liouville
system finding various excitations can be carried out by applying
the minimum principle.  

Using the fact that the ground state
energy in such a system corresponds to the absolute minimum of its
energy functional, we can prove rigorously an inequality
\eqn\VTInequality{m^2_{T,0} < m^2_{V,0},}
between the masses for the tensor and the vector ground states. This
inequality follows immediately from the fact that the potential
function for vector modes, $w_V(x)$, is strictly positive over the
range $0< x<\infty$. (See Appendix B.)

We next apply a variational approach to obtain upper bounds for the
respective ground state masses. Simple trial  wave functions 
satisfying the respective tensor and vector boundary conditions at
$x=0$ and at $x\rightarrow
\infty$ are $\bar{\phi}_{T}(x)=(x+1)^{-2}$ and $\bar{\phi}_{V}(x)
=(x+1)^{-{3}}\sqrt{x(x+2)}$. One
readily obtains the following  simple upper bounds for the ground state masses:
\eqn\TVSimpleBounds{  m^2_{T,0}\leq 12,\quad\quad {\rm and} 
\quad \quad m^2_{V,0}\leq 20.}
These values compare very well with the precise numerical answers:
$m^2_{T,0}=11.588$ and $m^2_{V,0}=18.676$.  One can easily improve
these already impressive bounds by using a better  variational ansatz.
For instance, using $\phi_{T}(x) = (x+1/x +\lambda) \bar{\phi}_{T}(x)$ and
$\phi_{V}(x) = (x+1/x + \lambda) \bar{\phi}_{V}(x)$, with $\lambda$ as 
variational parameter, we obtain
\eqn\BestBoundT{m_{T,0}^2\leq 11.588, \quad {\rm for} \quad 
\lambda=2.0857\>,}
\eqn\BestBoundV{ m_{V,0}^2\leq 18.687,
\quad {\rm for}
\quad \lambda=1.1729\>.}

The situation for scalar modes is slightly more
involved. Nevertheless, an analogous analysis can also be carried out,
as explained in Appendix B. For scalar wave-functions,
$\phi_S(x)\equiv (r^4-1) q_S(r)/r$, the simplest choice for
ground-state trial function consistent with boundary conditions at
$x=0$ and $x\rightarrow \infty$ is ${\bar{\phi}_{S}(x)={\rm
constant},}$ which leads to an upper bound for the scalar ground-state
mass:
\eqn\SSimpleBound{  m^2_{S,0}\leq 5.5213.}
Observe immediately that this upper bound is already very close to the 
exact numerical result:
$m^2_{S,0}= 5.4573$. An improved
variational bound can obviously be obtained, as was done for both the 
tensor and the vector
ground states. 
It is also important  to note  that this  simple variational
upper bound for the scalar ground state
is already much lower
 than the  mass for the tensor ground state. 

\subsec{\bf WKB  Estimates For High-Mass Excitations}
We next focus on high-mass excitations by carrying out a WKB analysis.
We begin by first converting our Sturm-Liouville type equations into
 ``radial" type Schroedinger equations,
\eqn\Radial{\big( -{ \hbar^2} {d^2\over dx^2} + V(x; {
m^2})\big )
\psi(x) = E
\psi(x),}
where $E\rightarrow 0^{-}$.  In keeping with the standard practice of
WKB methods we have also introduced a formal expansion parameter
$\hbar$, which at the end of the calculation we set back to $\hbar =
1$.  The relations between the new set of wave functions, $\psi_V(x)$,
$\psi_T(x)$ and $\psi_S(x)$, and the previous set, $\phi_V(x)$,
$\phi_V(x)$ and $\phi_S(x)$, are given in Appendix C. Note that, in
this approach, eigenvalue parameter $m^2$ appears as a parameter in an
effective potential, and it controls the ``strength" of  ``central
force" necessary in forming ``zero-energy" bound states. For all three
cases, when $m^2$ becomes large, it can be shown that each potential
approaches a smooth large-$m$ limit:
\eqn\CentralPot{\eqalign{V(x;m^2)&\rightarrow m^2 V_0(x) + 0(1),\cr
V_0(x)& = - {1\over 4x (x+1)(x+2) }.\cr}} By dividing Eq. \Radial\ by
$m^2$, we observe that, in the large $m$-limit, $\hbar$ and $m$ appear
naturally in a simple ratio, ${\hbar/ m}$. Therefore, a large-$m$
limit is formally the same as a``small"-$\hbar$ limit. (Strictly
speaking,  this is not the case for our scalar equation.) Or less
formally stated, as we observe in our numerical solutions, the mass
eigenvalues $m^2_n$ increase monotonically with the number of zeros
$n$ in the wave function so that the large eigenvalues expansion is
also a short distance limit, as expected.

For bound-state problems, the  WKB consistency condition for
eigenvalues $m = m_n$  for $n=0, 1,2,\cdots\cdots,$ is  given as an
expansion in $\hbar$,
\eqn\WKB{(n+{1\over 2}) \> {\pi\hbar } = I_0(m) + {\hbar^2}\>\>I_2(m) +  
\>{\hbar^2}\>\>I_4(m) +
\cdots\cdots \quad,} 
where the coefficients $I_{2k}(m)$ are expressed as an integral
between two classical turning points over integrand expressed in term
of the potential. Since the leading behavior of the integrals are
$I_{2k}(m) \sim m^{1- 2k}$ at large $m$, we obtain an asymptotic
expansion in $1/m$ at $\hbar = 1$.  Systematically expanding in
${1/m}$, the WKB condition takes the form:
\eqn\NewWKB{(n+{1\over 2}) \> {\pi} = s_0\>  m+ s_1\> + { s_2\over m}+ 
\cdots,}
where coefficients $s_i$ are easily calculable.

We have shown in Appendix D that, in order to obtain the first two
coefficients, $s_0$ and $s_1$, it is sufficient to consider only the
leading order WKB contribution,
\eqn\WKBZero{ I_0(m) =  \int_{x_L}^{x_R} d x 
\sqrt{ -V(x; {m^2})- {1+4\epsilon\over 4 x^2} },
 } where $x_L(m)$ and $x_R(m)$ are two classical turning points with
$\epsilon\rightarrow 0^{+}$. The shift from $V(x;m^2)$ to $V(x; m^2) +
{1\over 4 x^2} $, first discussed by K. E. Langer~\langer, is commonly
known as the ``Langer correction".  It incorporates the correction at
the singular boundary when there is a centrifugal term, $l(l+1)/x^2$,
in a radial wave equation. Furthermore we have added an $\epsilon$
term, $ {1/4 x^2}\rightarrow {(1+4\epsilon) / 4 x^2} $, as a simple
procedure for properly implementing the threshold behavior for
$\psi(x)$ at $x=0$. (See Appendix D.)

>From Eq. \WKBZero, one notes that $I_{0}(m)$ can be a non-trivial
function of ${m}$ through its dependence in $V(x; m^2)$ and on the two
classical turning points, $x_L(m)$ and $x_R(m)$. Since $V\sim m^2$ for
$m$ large, it follows that $I_0(m)= m\big (I_{0,0} +m^{-1} \> I_{0,1}
+ m^{-2} \> I_{0,2}+ \cdots\big )$ at large m. Consequently
$s_0=I_{0,0}$, and $s_1 = I_{0,1}$. (To obtain
$s_2$, one  will also need to calculate $I_2$.)
Truncating Eq.\NewWKB\ to this order and solving for $m^2$, one obtains  the
following  expansion  for the mass spectrum,\minahan\
\eqn\WKBMass{m_n^2/\mu^2 \>\>  \simeq \>\>     n^2 \> + \>  \delta \> \> n \>    +
\>  \gamma\> ,} 
where $\mu^2$,
$\delta$, and $\gamma$ are related to $s_0$, $s_1$ and $s_2$ by $\mu =
\pi/ s_0$ , $\delta = 1 - 2 s_1/ \pi $, and $\gamma=\delta^2/4 - 
2s_0s_2 / \pi^2$.
%New delta = delta + 2
%New gamma = gamma + 1 + delta
 
We have carried out this WKB analysis for all three cases, and have obtained 
the following
interesting results.
First, we have found that the leading coefficient, $\mu^2$,  is 
{ universal},
{\it i.e.},
\eqn\LeadingCoeff{\mu^{-1} = ({2\over\pi})  \int_0^\infty { d x}
{\sqrt
{-V_0(x)}}=  {1\over 4}{ B({\textstyle 1\over 4}, {1\over 2})},}   
is the same  for tensor, vector,  and scalar modes.  Therefore, it is
also the same as that  for the dilaton
modes, with $\mu^2 \simeq
5.742\> .$
Second, for the term linear in $n$, simple integer coefficients  are
 obtained for both the
tensor and vector modes: 
\eqn\VTDelta{\delta_T=3 \quad\quad\quad \delta_V=4.} 
However, for  scalar modes, a more intricate calculation is needed. We find
 that 
\eqn\SDelta{\delta_S= {1\over
\pi}
\int_{0}^{u_r} {d u\over u^{3/2}}\Big( 1-{
\sqrt {{1-3u^2 -9u^3}}
\over {1+3u}}\Big) + {2\over \pi \sqrt{u_r}} + 1\simeq 3.018\>,}
where $u_r$ is the positive root of 
$9u_r^3 + 3 u_r ^2 -1 =0$, $u_r\simeq 0.392\>. $
Putting these together, we have
\eqn\VWKBMass{\eqalign
{m_T^2\simeq & 5.742\> ( n^2 + 3 n + \gamma_T),\cr m_V^2\simeq &
5.742\> ( n^2 + 4 n + \gamma_V),\cr m_S^2\simeq & 5.742\> ( n^2 +
3.018 n +
\gamma_S).\cr}}
In Figure 1, we show a comparison on a log-log plot
of our WKB prediction to our direct numerical results up to $n=9$ for
all three distinct modes. Interestingly, except for the lowest scalar
mass, our WKB results agree with the actual data extremely well, with
constants, $\gamma_T\simeq 2.011$, $\gamma_V\simeq 3.283$, and
$\gamma_S\simeq 1.25$ fitted to the asymptotic form at large
n. Higher order WKB results and other related work will be
reported elsewhere.

\ifig\fwkblog{Comparison of WKB approximation
with the levels computed in Table 1.}
{\epsfxsize4.25in\epsfbox{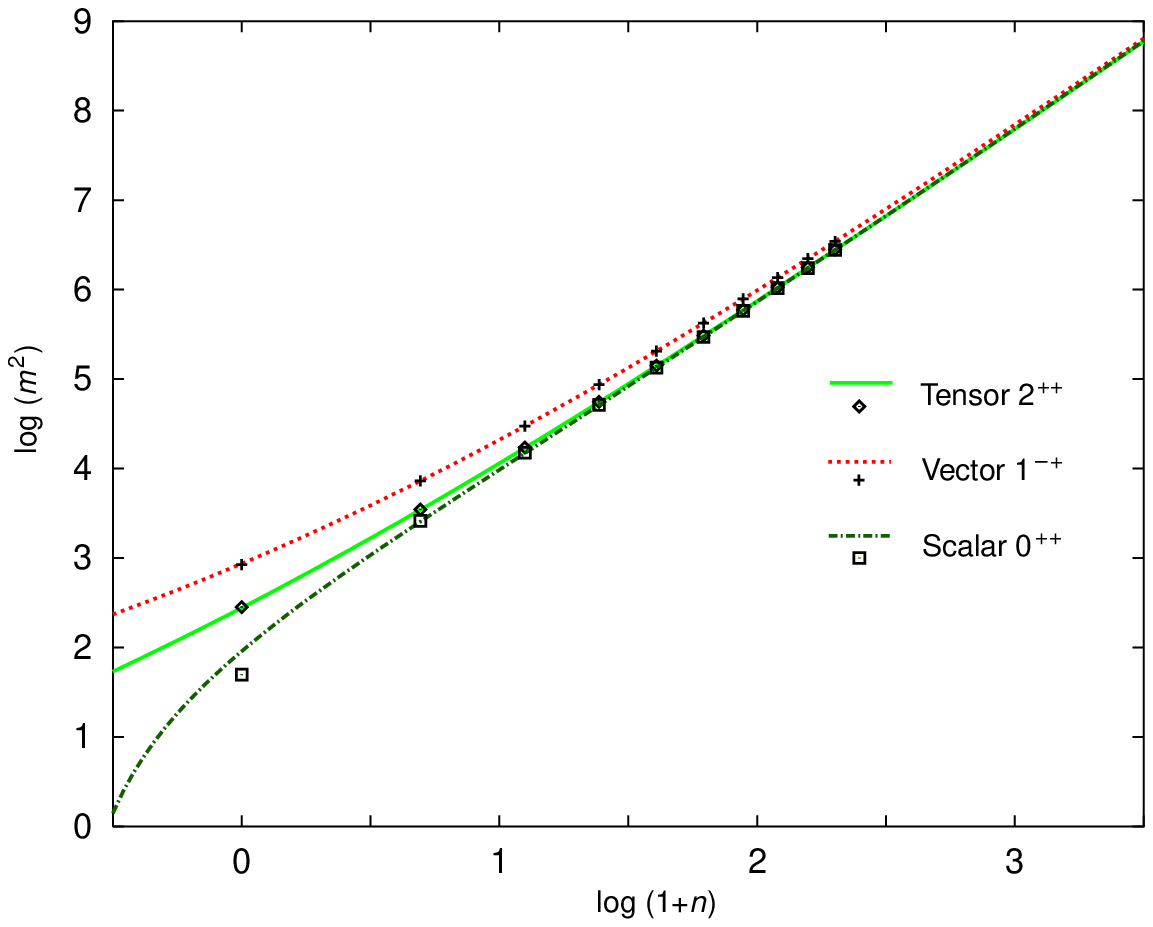}}

In spite of this impressive agreement between the low order WKB
calculations and the exact numerical results, a cautionary note
should be added. As pointed out in Ref~\csaki, the question of
convergence of  $1/m$-expansion for the logarithm of 
wave functions,   $\phi_V(x)$,
$\phi_V(x)$ and $\phi_S(x)$, remains to be addressed.

\subsec{\bf Heuristic Picture:}
It is possible to provide a qualitative  understanding for  both
the low-mass pattern,
 $m^2_{S,0}<
m^2_{T,0} < m^2_{V,0}$, and the asymptotic high-mass excitations by making 
use of a quantum
mechanical analogy. For $m$ large, Eq. \CentralPot\ formally corresponds to a 
``screened" attractive Coulomb potential, with a  ``bare" charge $Q$ at the
core,
\eqn\CoreCharge{Q=-{m^2\over 8}+ 0(1). }
More generally, we have shown in 
Appendix C that, for all three cases,  each 
potential,
$ V(x; m^2)$, behaves 
near $x\sim 0$ as
\eqn\SingularityZero{V(x; m^2) \simeq {L^2\over x^2} +
 {Q\over x} + 0(1).}
This naturally suggests that we identify $L^2=l(l+1)$ as angular momentum 
squared and $Q$ as
the unscreened bare charge at the core. We find that
$l_V=0$, and
$l_T=l_S=-{1\over 2}$.

Our boundary condition for $\psi$ at $x=0$ corresponds to choosing
$\psi(x)\sim x^{l+1}$ over $\sim x^{-l}$. For ${\rm Re}\>\> l \geq
-{1\over 2}$, this would correspond to always selecting the ``regular"
solutions. In order to avoid the degeneracy ambiguity for the case
where $l_T=l_S=-{1\over 2}$, we shall in what follows always assume
that $l_T=l_S= -{1\over 2}+ \epsilon$ and consider the limit $\epsilon
\rightarrow 0^+.$

When we include the ``Langer correction", $V\rightarrow V+ ({1\over
4}+\epsilon)({1\over x^2}) $, the angular momentum factor, $L^2=l(l+1)$, becomes
$
\tilde L^2 = (l+{1\over 2})^2 +\epsilon$.  For $x\rightarrow 0$, we find that
\eqn\LQ{\eqalign{\tilde L^2_V={1\over 4},\quad\quad &Q_V= -{m^2-6\over
8},\cr \tilde L^2_T= \epsilon, \quad\quad\> &Q_T= -{m^2-12\over 8},
\cr \tilde L^2_S=\epsilon, \quad\quad\> &Q_S=-{m^2-2+ 16 m^{-2}\over
8}\cr}.}  
At large $m^2$, \LQ\ leads to~\CoreCharge, in agreement with
Eq.~\CentralPot.  More generally, since $ V(x;m^2)\rightarrow
m^2V_0(x)$ for all three cases, this explains our finding that the
coefficient $\mu^2$ is universal.  The next order correction,
$\delta$, will depend on how $ V(x;m^2)$ deviates away from $m^2
V_0(x)$, both at $x\sim 0$ and at $x\sim \infty$. Indeed, as we show
in Appendix D that our results for $\delta_T$, $\delta_V$ and
$\delta_S$ follow directly after we have identified the leading
asymptotic behavior of $V$ at $x=0$ and at $x=\infty$.

Turning next to the low-mass spectrum. In this analog picture, the
vector ground state is more massive since it corresponds to a higher
angular momentum state when compared to the tensor ground state,
$\tilde L^2_V>\tilde L^2_S$, and therefore a stronger binding, (larger
$m^2$), is required in order to form a zero-energy bound state.  In
contrast, the tensor modes and the scalar modes have the same angular
momenta, $\tilde L^2_T=\tilde L^2_S=\epsilon>0$.  We find that the scalar
``potential-well" is ``deeper" than that for the tensor modes,
especially when $m^2$ is small. Indeed, whereas the Coulomb core for
tensor modes only become attractive when $m^2\geq 12$, it is always
attractive for the scalar modes, with minimum attraction at $m^2=4$.
Therefore, less binding is required in forming a zero-energy scalar
bound state, and it is consistent with the fact that the scalar ground
state is lighter than the tensor mode.

\newsec{Discussion}

We have computed the discrete spectrum of metric perturbations
in an $AdS^5$ black hole background.  We would like to end by briefly 
discussing our  results and making a few cautionary remarks on the
difficulties of the comparison of this strong coupling spectrum with
glueball masses as computed in lattice gauge theory or classified in bag
models.

 As stressed in Ref.  \jev, in order to obtain these discrete
modes, a non-singular boundary condition must be applied at the
horizon of the black hole and requiring normalizibility of wave-functions
alone is insufficient for such a purpose. A deeper understanding on
the physics of the boundary condition used would be highly
desirable. Our discussion has also been restricted to modes that are
singlets under the SO(6) symmetry and the ``finite temperature'' Kluza
Klein U(1) gauge group. As a guide to finding states that might survive the
dimensionally reduced limit, concentrating on states that
are non-singlets in
$SO(6) \times U(1)$ appears reasonable. 

Earlier comparisons for
$QCD_3$ have noted an apparent agreement between the ratio,
$m(0^{--})/m(0^{++}) = 1.45
\pm 0.03$, found in weak coupling lattice simulations and the ratio,
$m(0^{--})/m(0^{++}) \simeq  1.50$, found at strong coupling
from supergravity, using  the lowest state of the pseudoscalar mode
($B_{\mu\nu}$) and the lowest state of the dilaton ($\phi$). 
Our new lower mass
$0^{++}$ state will somewhat upset this quantitative correspondence. 
With  a much smaller value for the scalar mass, this
supergravity mass-ratio becomes 3.19 which is too large by a
factor of two. 

However, the earlier 
agreement was inexplicably good in any case.  From the point of
view of a 3-d Yang-Mills theory at weak coupling the lowest
$0^{++}$ state should couple to
both the operator $S = Tr[F^2_{\mu\nu}(x)]$ and the helicity-0
component of the energy-momentum tensor $T_{\mu \nu} =
Tr[F^2_{\mu\nu}(x)] - {1 \over 4}
\eta_{\mu \nu} S$, which are associated with the dilaton and the
scalar component of the metric fluctuations respectively. Indeed each of
these operators have contributions from the $\tau$-component of
$A_\mu$ along the $S^1$ compact direction, which from the view point
of 3-d Yang-Mills theory is an adjoint Higgs field. The state
associated with the Higgs mode must decouple at weak coupling if there
is a pure Yang-Mills fixed point.

A second feature that has previously been noted is the similarity in
the level spacing of the excitation spectrum of supergravity states
and the spectrum of the excited glueball spectrum in lattice
simulations.  In our view, this is even more problematic. For one thing,
we have so far  ignored Kluza-Klein excitations in the compact $S^1$
direction that have a level spacing given by  $M_{KK} = 1/R \equiv 2 b$, where
$R$ is the size of the compactified circle and the parameter $b$ so defined has
units of mass. 
This spacing is of the same order as the mass gap for the lowest
$0^{++}$ state: $m(0^{++}) \simeq  \sqrt{5.457}~b \simeq 2.3~ b$. This
suggests that the proper identification of the excitation spectrum of
the supergravity modes with states in Yang-Mills theory is
non-trivial.  In quarkless QCD, all mass splittings are  controlled by
a single mass scale,
$\Lambda_{QCD}$ in 4-d or $g^2_3 N$ in 3-d, which one may equally well
take as the string tension, $\sigma$. In the strong coupling limit on
the lattice, this scale is replaced by the UV cut-off (or inverse
lattice spacing $a^{-1}$). In these units both the lattice glueball
masses and the string tension diverge as $log(g^2N)$ or $log(a g^2_3 N
)$, whereas in the supergravity scenario at strong coupling the
putative 3-d ``glueball'' masses go to  constants while AdS string
tensions diverges as
\eqn\tension{\sigma = {{\sqrt{4 \pi g_s N}\; b^2} \over {2 \pi}}\;,} 
in leading order in strong coupling.  It may be that  lattice glueball
excitations are  transverse stringy fluctuations that
would not survive in the the supergravity approximation. In that
case, from a
lattice view point, there should be {\it no} infinite tower of
``glueball'' states. The supergravity spectrum describes radial (5-th
dimensional)  excitations that either do not survive the weak coupling
finite N limit or are mixed with stringy states. It is difficult to see
what should be the precise comparison between these two sets of excitations
since they have quite different physical features. 
We
consider the understanding of the correspondence of this tower of
states with Yang-Mills theory at the boundary to be an important
unsolved issue.

In spite of these reservations, we believe that a comparison between the supergravity results and lattice as well as bag model results does offer  opportunities  to uncover phenomenologically
interesting features. For instance, in the mass relations for the lowest
states as a function of their quantum numbers, we do see a pattern
reminiscent of lattice glueball spectra.  Indeed the pattern,
$$m(0^{++}) < m({\tilde 0}^{++})=m(2^{++}) < m(1^{-+})$$
 is generic to
both lattice calculations in 3-d and bag calculations, which are in
excellent agreement with each other.  We find that the classification of
states with low-dimension color singlet fields in a bag
picture for glueballs, {\it i.e.}, in terms of  ``valence-glue'',  follows the same
pattern as the expected in the AdS/CFT correspondence.  A fuller
understanding of these relationships is certainly worth pursuing. 
First
it is evident from our analysis above that all  128 lowest
bosonic modes for the supergraviton multiplet that are $SO(6)
\times U(1)$ singlets need to be found. For 3-d Yang-Mills, it may be
also more  fruitful to make comparisons with high temperature lattice
$QCD_4$ results, where the constant KK adjoint field in the
compact $\tau$ direction still remains.

Our current exercise needs to be extended to schemes
for 4-d QCD such as the finite temperature versions of $AdS^7 \times
S^4$. As has been suggested elsewhere, the goal may be to find that
background metric that has the phenomenologically best strong coupling
limit. This can then provide an improved framework for efforts to find
the appropriate formulation of a QCD string and for addressing the
question of Pomeron intercept in QCD.~\tanO\tanT~ In addition, a more
thorough analysis of the complete set of spin-parity states for the
entire bosonic supergravity multiplet and its extension to 4-d
Yang-Mills models is also  worthwhile. Results on
these  computations will be reported in a future publication.

\vskip20pt

{\noindent\bf Acknowledgments:} We would like to acknowledge useful
conversations with  R. Jaffe, A. Jevicki, D. Lowe, J. M. Maldacena, H.
Ooguri, and others. This work was supported in part by the Department
of Energy under Contract No. DE-FG02/19ER40688-(Task A).

\vskip20pt
{\noindent \bf Note added:} A brief version of our results was presented in 
 talks by R. Brower
at Lattice '99 (Pisa, July 1999) and also at ``QCD and Multiparticle
 Production--ISMD99", (Brown,
August 1999). While this paper was being finalized, a paper~\cm\  
appeared, which has a considerable overlap with our results.

\appendix{A}{Field Equation for Spin-0 Perturbation:}

Let us start with the gauge where the nonzero components of the spin-0
perturbation are
\eqn\sixtqq{\eqalign{h_{11}&=h_{22}=q_1(r)e^{-mx_3}\cr
h_{\tau\tau}&=-2q_1(r){f(r)\over r^2}e^{-mx_3}+q_2(r)e^{-mx_3}\cr
h_{rr}&=q_3(r)e^{-mx_3}\cr }}

The Einstein equations \thir\ give
\eqn\sevent{q_3= - {1\over (r^4-1)^2(3r^4-1)} 
[2 r^8q_2(r)-r^5(r^4-1)q_{2,r}(r)+ 4(r^4-1)q_1(r)] e^{-mx_3} }
and one first order and one second order equation in $q_1$ and $q_2$:
\eqn\oneq{\eqalign{4r(3r^4-1) q_{1,r}
+3r^5(r^4+1)q_{2,r}&
-8 (3r^4+1)q_1\cr
-[6r^4(r^4-1)&+m^2{r^6(3r^4-1)\over r^4-1}]q_2
=0\cr}}
\eqn\threeone{\eqalign{
3r^2(r^8-1)q_{1,rr}
+r[3(r^4+3)&(r^4+1)-4m^2r^2]q_{1,r}\cr
+[-12(r^4+1)^2&+m^2r^2(3r^4-1)]q_1
-m^2r^6[m^2{r^2\over r^4-1}+3]q_2=0\cr }}
We can find $q_2$ from \threeone\ and substitute it into \oneq\ to obtain a
third order equation for $q_1$:
\eqn\twenty{\eqalign{
r^3(r^4-1)^2&[m^2r^2+3(r^4-1)]q_{1,rrr}\cr
+
r^2(r^4-1)&[-m^4r^4+2m^2r^2(r^4+5)+3r^2(r^4-1)^2(3r^4+5)]q_{1,rr}\cr
+
r[-4m^4r^4+&m^2r^2(23-6r^4-r^8)-3(r^4-1)^2(3r^4+13)]q_{1,r}\cr+
[-m^6r^6+m^4&r^4(r^4+3)-2m^2r^2(13+2r^4+r^8)+48(r^4-1)^2]q_1=0\cr}}
It may appear strange that we have obtained a third order system (which needs
three constants of integration) instead of a second order system as in the
spin-2 and spin-1 cases above. The source of this extra freedom is that the
gauge
\el\ there is a one parameter freedom left in the present ansatz:
\eqn\twone{\eqalign{x_3\rightarrow& x_3+\epsilon a(r)e^{-mx_3}\cr
r\rightarrow& r+\epsilon {ma(r)\over rf(r)}e^{-mx_3}\cr} }
with
\eqn\twtwo{a(r)= \exp[\int {dr\over r} (2+{k^2\over f})]}
This freedom gives a change in $q_1$ proportional to
\eqn\twthree{q_1^{gauge}\equiv
-{2-mr^2(1-r^2)^{m^2/4}\over (1+r^2)^{m^2\over 4}} }
Substituting 
\eqn\twfour{q_1=q_1^{gauge}\tilde q_1}
we get a second order equation 
\eqn\twfive{\eqalign{
r^2(r^4-1)^2&[-m^2r^2-3(r^4-1)]\tilde q_{1,rr}\cr
+
r(r^4-1)&[-2m^4r^4-m^2r^2(17r^4-5)-3(r^4-1)(9r^4-1)]\tilde q_{1,r}\cr
-[m^6r^6+4m^4&r^4(3r^4-1)+4m^2r^2(5r^4-1)(2r^4-1)+3(r^4-1)^2(15r^4+1)]\tilde q_1
=0\cr }}

A similar third order system of equations arises in the study of polar
perturbations of the Schwarzschild metric, in the gauge where the metric is
constrained to be diagonal. One discovers a combination of fields that
`remarkably' satisfy  a second order equation, the `Zerilli equation'
~\chandra. But we can see that such a reduction is to be
expected because here again there is a one parameter family of diffeomorphisms
that preserves the gauge condition.  Consider a metric of the form
\eqn\fone{ds^2=k(r) dt^2+k^{-1}(r)dr^2+r^2(d\theta^2+\sin^2\theta d\phi^2)}
Then if we are looking at perturbations with frequency $\omega$ in time $t$,
then we have the following  diffeomorphisms which leaves the metric within the
chosen gauge (off diagonal components zero):
\eqn\ftwo{\eqalign{\theta\rightarrow & \theta+ h(r){\partial\over
\partial\theta}P_l(\theta)e^{\omega t}\cr r\rightarrow & r-r^2({h(r)\over
r^2})_{,r}P_l(\theta)e^{\omega t}\cr t\rightarrow & t-\omega
h(r)P_l(\theta)e^{\omega t}\cr}} where for infinitesimal $\epsilon$  
\eqn\fnine{h(r)=\epsilon rk(r)^{1/2} . }
($P_l(\theta)$ is the associated Legendre polynomial in $\cos\theta$.) We can
use this solution to reduce the third order system to a second order equation.

But for the Schwarzschild metric one can choose a different gauge, where such a
residual gauge freedom does not exist. Such a gauge was used in \rw . We
 thus look for a gauge in our
problem which would also have no residual gauge symmetry, and so 
directly yield
a second order equation.

Consider the gauge where the nonzero components of the perturbation are
\eqn\sixt{\eqalign{h_{\tau\tau}&=q_1(r)e^{-mx_3}\cr
h_{rr}&=q_2(r)e^{-mx_3}\cr
h_{r3}=h_{3r}&=q_3(r)e^{-mx_3}\cr }}
The Einstein equations \thir\ determine $q_2$ and $q_1$ in terms of $q_3$:
\eqn\thone{q_2=-{r^5(r^4-1)q_{1,r}+2r^8 q_1\over (r^4-1)^2(3r^4-1)} }
\eqn\thtwo{q_1={2(r^4-1)\over mr^5[m^2r^2+3(r^4-1)]}
[3r(r^8-1)q_{3,r}+(3(r^4+1)(3r^4+1)+m^2r^2(3r^4-1))q_3]}
and give a second order equation for:
\eqn\ththree{\eqalign{r^2(r^4-1)^2(3(r^4-1)+&m^2r^2)q_{3,rr}+r(r^4-1)[3(r^4-1)
(5r^4+3)+m^2r^2(7r^4+5)]q_{3,r}\cr
+[9(r^4-1)^3+&2m^2r^2(3+2r^4+3r^8)+m^4r^4(r^4-1)]q_3=0.\cr}}
This is Eq. \thtreeq, where we have re-labeled $q_3$ by $q_S$.
This second order equation is related to the second order equation \twfive\ by
the transformation
\eqn\thsix{\tilde q_1= {(1+r^2)^{m^2/4}\over r^2(1-r^2)^{m^2/4}}q_S}

\appendix{B}{Sturm-Liouville Approach and Variational Analysis:}

Eq. \SL\ is in  the standard Sturm-Liouville form ,
\eqn\BSLproblem{\Big(-{d\over dx}\tau(x) {d\over dx} \> + w(x)
\>\Big) 
\phi_n(x) = m^2_n
\sigma(x) 
\phi_n(x),}
where $\tau(x)=x(x+1)(x+2)$ and $ \sigma(x)={1\over 4}$.
The  set of eigenfunctions, $\{\phi_n\}$, provides a complete orthonormal 
basis: ${<\phi_n|\phi_m>\equiv \int_0^{\infty} dx \sigma(x)
\phi_n(x)\phi_m(x) = \delta_{m,n}}$. 
It is well-known that solving for this set of eigenfunctions and their
corresponding eigenvalues is equivalent to finding stationery points
of the following "energy functional",
\eqn\EnergyFunctional{\Gamma[\phi]\equiv {\int_0^{\infty}
dx[\tau(x){\phi'(x)}^2 
 + w(x) \phi(x)^2]\over  {\int_0^{\infty} dx  \sigma(x)
\phi(x)^2}.}} At each stationery point,
$\phi_n$, one has 
${\Gamma[\phi_n] = m^2_n.}$
Since both $\tau(x)$ and $w(x)$ are positive, $\Gamma[\phi]$ is bounded
from  below and is
in fact  positive. In particular, the square of the mass for the lowest
 state,
$m^2_0=\Gamma[\phi_0]$,  is the absolute minimum of
$\Gamma[\phi], $ and $m^2_0>0.$

Let us denote $\Gamma_{\{T/V\}}[\phi]$ as the energy functional for the tensor
and vector modes respectively. We shall next apply this formalism to obtain a
simple and yet  important  mass inequality: $$m^2_{V,0}> m^2_{T,0}.$$
Denote the the lowest tensor and vector masses by $m_{V,0}$ and $m_{T,0}$ 
respectively. Since each
corresponds to the absolute minimum of its  energy functional, it follows 
that 
\eqn\UpperBounds{0< m^2_{T,0}=\Gamma_{T}[\phi^{T}_0]\leq  \Gamma_{T}[\phi];
 \hskip40pt
0< m^2_{V,0}= \Gamma_{V}[\phi^{V}_0]\leq 
\Gamma_{V}[\phi].}
Alternatively, $m^2_{V,0}$ can also be expressed  as 
${m^2_{V,0}=  <w_V>_0+ \Gamma_{T}[\phi^{V}_0],}$
where 
$${<w_V>_0\equiv {\int_0^{\infty} dx w_V(x) 
{\phi^V_0(x)}^2 \over \int_0^{\infty} dx 
\sigma(x){\phi^V_0(x)}^2}.}$$
Note that $w_V(x) = 1/(x(x+1)(x+2))$ is a positive function for
$0\leq x<\infty$;  it follows that
$<w_V>_0$ is positive, and
\eqn\BoundOne{m^2_{V,0}> \Gamma_{T}[\phi^{V}_0].}
On the other hand, since $m^2_{T,0}$ is the absolute minimum of $\Gamma_T$, 
we also have the
following inequality,
\eqn\BoundTwo{\Gamma_{T}[\phi^{V}_0]> \Gamma_{T}[\phi^{T}_0]=m^2_{T,0}.}
>From \BoundOne\ and \BoundTwo, the mass inequality between vector and
 tensor ground states, 
${m^2_{V,0}> m^2_{T,0},}$ Eq.
\VTInequality, follows.

It is also worth  pointing out that, in re-deriving these eigenvalue 
equations from the energy
functional, it is  necessary to impose boundary conditions
\eqn\Bdry{\tau(x) \phi(x) \phi'(x) \rightarrow 0,}
both for  $x \rightarrow 0$ and  $x\rightarrow \infty$.  This states more 
clearly what 
boundary condition is to  be used at the horizon of the black hole. In 
particular,  this rules out
the solution where $\phi(x)\sim \log x$ at $x=0$. 
To be more precise, by examining our  differential equations, Eq. \SL, 
both at $x=0$ and at
$x\rightarrow
\infty$, Eq.
\Bdry\ requires that tensor and vector wave-functions must behave as 
\eqn\BOne{\eqalign{\phi_{T}(x)\rightarrow x^{-2}, \quad\quad\quad 
\phi_{V}(x)\rightarrow
x^{-2},\quad\quad \quad&x\rightarrow \infty,\cr
\phi_{T}(x)\rightarrow constant, \quad \phi_{V}(x)\rightarrow x^{1/2},
\quad
&x\rightarrow 0.\cr}}

 We next address the question of  scalar modes. Let us first convert Eq. 
\fift\ into 
\eqn\ScalarDE{{\hat L}_S \phi_S(x) \equiv \Big(-\big( {d\over dx}
\tau_S(x;m^2) {d\over dx}\big)  \>   +
w_S(x;m^2)\Big )\phi_S(x)\>=\> 0,} 
where 
\eqn\TauS{\tau_S(x;m^2)=  {x(x+2)\over [3x(x+2) + m^2 (x+1)]},
\quad\quad w_S(x;m^2)= { m^2  [ (3x^2+6x+4) -
m^2 (x+1)]\over 4(x+1)[3x(x +2) + m^2(x+1)]^2}.} Here, $\phi_S(x)$ is
related to $q_S(r)$ in \fift\ by $\phi_S(x)= {x(x+2)\over \sqrt {x+1}}
q_S(r)$.  Since $ m^2$ enters into the differential equation in a
rather non-trivial fashion, we are unable to directly convert this
into a Sturm-Liouville problem with $m^2$ as eigenvalues.  Let us
consider instead the following eigenvalue problem:
\eqn\SLScalr{{\hat L}_S \phi_n(x) = E_n \sigma_S (x) \phi_n(x),}
where $ E_0 < E_1<E_2< \cdots,$ and $\sigma_S (x)$ is at this point
unspecified. This in turn can be obtained via a Sturm-Liouville type
variational approach, i.e., finding stationery points of the following
energy functional:
\eqn\ScalarFunctional{\Gamma_S[\phi] \equiv {\int_0^{\infty} dx 
\big (\tau_S(x;m^2) |\phi(x)' |^2 +
w_S(x;m^2) |\phi(x)|^2\big)\over \int_0^{\infty} dx 
{\sigma_S (x) |\phi(x) |^2.}}.}
Note that, here, $m^2$ is to be treated as a {\bf parameter.} 
To have a
properly defined stationery conditions, we require
$\tau_S(x;m^2)\phi(x)\phi(x)'=0$, 
both at $x=0$ and $x\rightarrow \infty.$ This, in turn, fixes our boundary 
conditions. A
careful examination indicates that the desired boundary conditions for 
$\phi_S(x)$ are
\eqn\BdryS{\phi_S(x)\rightarrow constant,}
for both $x\rightarrow \infty$ and $x\rightarrow 0$.

Since $w_S(x;m^2)$ is no longer positive, neither is $\Gamma_S[\phi]$.
Nevertheless, $\Gamma_S[\phi]$ is still bounded from below since
$w_S(x;m^2)$ is. As such, we identify $E_n^2(m^2) =\Gamma_S[\phi_n]$,
in increasing order, at each local minimum, $\phi_n$. In particular,
$E_0(m^2)$ is the absolute minimum of $\Gamma_S$, although it no
longer is positive in general.  This ``generalized Sturm-Liouville
problem" reduces to our scalar problem if a { zero-energy} solution
exists. To be more precise, one needs to answer the following
question: ``{For what values of $m_j^2$, $j=0, 1, 2, \cdots$, will
there be a zero-energy solution where $E^2_j(m^2_j)=0$ ?}"  As a
minimum problem, we are searching for $m_0^2$ for which the absolute
minimum of $\Gamma_S(\phi)$ vanishes, {\it i.e.}, ${\rm min}\>
\Gamma_S[\phi;m^2_0]=0$. Turning this into a variational problem, we require
that, for $m^2>m^2_0$, 
\eqn\SLBoundScalar{0=E_0\leq {\int_0^{\infty} dx \big (\tau_S(x;m^2) 
|\phi(x)' |^2 + w_S(x;m^2) |\phi(x)|^2\big )\over
\int_0^{\infty} dx \sigma_S(x) |\phi(x)|^2},}
for suitably chosen trial wave-functions. For each choice of
$\phi(x)$, the value of $m^2$ at which the integral on the right
vanishes provides an upper bound for $m^2_0$. In arriving at our bound
for the scalar ground-state mass, Eq. \SSimpleBound, the trial
function ${\bar{\phi}_{S}(x)={\rm constant},}$ was used.

\appendix{C}{Radial Schroedinger Representation and Effective Potentials:}

In order to bring  our ODE, Eq. \SL,  into the radial Schroedinger form, 
Eq. \Radial, 
we perform the following  transformation:
\eqn\BSch{\eqalign{\psi_T(x) = & {\big({x(x+1)(x+2)}
\big)^{1\over 2}}\phi_T(x),\cr
	\psi_V(x) = & {\big({x(x+1)(x+2)}\big)^{1\over 2}}
\phi_V(x),\cr
	\psi_S(x) = & {\big({ m^2 x(x+2)\over  3x(x+2)+
{m^2(x+1)}}\big)^{{{1\over
2}}}}\phi_S(x).\cr}} 
The resulting effective potentials for all three cases are:
\eqn\BPotential{\eqalign{
V_{T}(x;m^2) = & m^2 V_0(x) +  { 3 
(x^2+2x+2)\over 4x (x+1)^2(x+2)} 
-{1\over x^2 (x+1)^2(x+2)^2},\cr
	V_{V}(x; m^2) = & m^2 V_0(x)+{ 3 
(x^2+2x+2)\over 4x (x+1)^2(x+2)},\cr
	V_{S}(x;m^2) = & m^2 V_0(x) +{m^2(3x^2 + 6x +2 )\over 2 x(x+1) 
(x+2)[m^2 (x+1)
 + 3 x (x+2)]  } + \Delta V_S,\cr}}
where  a rather involved expression has to be added  for the scalar potential,
\eqn\BPotentSS{\Delta V_S = - \big( {((x+1)^2+1)^2 + 4 x (x+2)\over
 4 x^2
(x+2) ^2}\big)\Big({
m^4+ {4m^2 x (x+1) (x+2) (3 (x+1)^2 +7)\over 
(((x+1)^2+1)^2 + 4 x (x+2))}
\over [m^2 (x+1) +   {3 x (x+2)} ]^2  }\Big),     }
Let us briefly comment on these effective potentials,
$\tilde V(x;m^2)= V(x;m^2) + (1/4+\epsilon)(1/ x^2)$,  in various limits,

\noindent (a) $m^2\rightarrow \infty$: For all three
effective potentials,
\eqn\LM{\tilde V(x;m^2) =  m^2 V_0(x) +0(1).}
(b)  $x\rightarrow 0$: In going from $V$ to $\tilde V$, 
for $x\sim 0$, we have
${\tilde V(x; m^2) \simeq {\tilde L^2/x^2} + {Q/x} + 0(1),}$
where  $ \tilde L^2 =(l+{1\over 2})^2+\epsilon$. Those $\tilde L^2$ 
and $Q$ values given in Eq. \LQ\  follow from
\eqn\SRadiusTS{\eqalign{
\tilde V_{T}(x;m^2)=&{\epsilon\over  x^2} -{ m^2 - 12\over 8x}+ 
0(x^{0}),\cr
\tilde V_{V}(x;m^2)=&{1\over 4 x^2}- { m^2 - 6\over 8x }+ 0(x^{0}),\cr
\tilde V_{S}(x;m^2)=&{\epsilon\over  x^2}-{ (m^2 -2+{16\over m^2})\over 8x}
+ 0(x^{0}).
\cr}}
(c) {$x\rightarrow \infty$}:  
\eqn\LRadius{\eqalign{ 
\tilde V_{T}(x;m^2)=& {1\over  x^2} - {m^2+6\over x^3} 
+ 0(x^{-4}),\cr
\tilde V_{V}(x;m^2)= &{1\over  x^2} 
- {m^2+6\over x^3} + 0(x^{-4}),\cr
\tilde V_{S}(x;m^2)= &{1\over 4 x^2}- {m^2\over12 x^3}
+ 0(x^{-4}).\cr}} From these, one concludes that each ``Coulomb" bare
charge, $Q$, is always screened at large distance; an effective
potential vanishing as $1/x^2$ at large distance is overall
``neutral".

\appendix{D}{WKB Treatment:}

In a standard WKB treatment for a one-dimensional quantum mechanics
problem, one relies on having a small parameter, ``$\hbar$", which
allows a systematically expansion for the logarithm of the
wave-function, $\phi(x) \sim \exp[\pm { i\over \hbar} \int^x dx
\sqrt{E-V(x)} + \cdots]$.  For a problem involving a spherically
symmetric potential, $V(x)$, one has a one-dimensional radial
Schroedinger equation involving an effective potential, {\it e.g.}, Eq.
\Radial, where  $V(x) \rightarrow V_{eff}(x) = V(x) +  l(l+1)\hbar^2/x^2$. 
In solving the radial equation, normalizability requires that the
wave-function $\psi \sim x^{l+1}$ as $x\rightarrow 0^+$, which also corresponds to our "non-singular" boundary condition at the black hole horizon. Note that
this exact threshold behavior, $\psi(x)\sim e^{(l+1) \log x},$ would
normally appear only in the next to leading order in a WKB expansion.

It is indeed possible to arrange so that the threshold behavior is properly 
implemented in the leading order by
working with a ``rapidity" variable, $y\equiv \log x$,
where $y$ is now unrestricted, 
$-\infty < y <
\infty$, and the radial Schroedinger Equation becomes
$\big (-{d^2/ dy ^2} + U(y) \big) \tilde \psi(y) =
\tilde E  \tilde \psi(y),  $
where 
$\tilde\psi(y) = e^{-{y\over 2}} \psi(x)$, $\tilde E \equiv E e^{2y}$, and
$$U(y)\equiv  e^{2y}\tilde V_{eff} (x) \equiv x^2 
\big ( V_{eff}(x)+ {\hbar^2\over 4 x^2} \big )=x^2
\big (  V(x) +{(l+{1\over 2})^2\hbar^2\over  x^2} \big ),$$
Note that, for $U(y)$ and $\tilde V_{eff}(x)$, the combination 
$l(l+1){\hbar}^2$ has become
$(l+{1\over 2})^2{\hbar}^2$, which correlates precisely with the appropriate 
threshold behavior for $\tilde\psi \sim
x^{l+{1\over 2}}$. For an attractive potential
$V(x)$ which is less singular than $1\over x^2$ at the origin, this new
effective potential
$U(y)$ will have an attractive well, with $U(y)\rightarrow  
(l+{1\over 2})^2{\hbar}^2$ when $y\rightarrow -\infty$,
whose bound state energies can be calculated via a standard WKB approximation.
For instance, the leading order WKB contribution is 
$ (n + {1\over 2}) {\pi}\hbar   = \int_{y_L}^{y_R} d y \sqrt {\tilde E-U(y)}, $
where $n=0,1,2,\cdots$, $e^{y_L}=x_L$ and $e^{y_R} = x_R$.

Returning  to our problem; with $E\rightarrow 0^-$, and $1/m$ playing
 the role of $\hbar$, we arrive at 
\eqn\WKBAppen{ I_0(m) = {1\over m} \int_{y_L(m) }^{y_R(m)} d y 
\sqrt{ -U(y; m)}. } 
By changing variable from $y$ back to $x$, Eq. \WKBAppen\ can also be
written as Eq.\WKBZero\ where where $\tilde V_{eff} (x; m)$ enters.  Let
us turn next to the expansion of $I_0(m)$ in $1\over m$, $I_0(m)
\simeq I_{0,0} + {1\over m} I_{0,1} + {1\over m^2} I_{0,2} + \cdots$.
Before proceeding, we separate the integral in Eq.  \WKBAppen\ , into
two,
\eqn\LR{I_0(m) = I_L(m) + I_R(m) ={1\over m} \int_{y_L(m)}^{0} d y 
\sqrt {-U(y;m)} +
{1\over m}\int_{0}^{y_R(m)} d y \sqrt {-U(y;m)}.}  This separation at
$y=0$ is of course arbitrary; our final result will be independent of
this choice.  It is convenient to separate $U(y;m)$ into two pieces,
$U(y;m)=m^2\big ( U_0(y) + m^{-2} U_1(y;m^2)\big)$, which will
be used in Eq. \LR.  To expand in $1/ m$, one needs to know how
$U_0$ and $U_1$ behave as $y\rightarrow \pm \infty$.

Let us  begin by treating the case of tensor modes with reasonable amount of 
details. Consider $I_L(m)$ and let us
examine the lower integration limit, $y_L(m)$. With both $\epsilon \sim  0^+$
and ${1/ m}\rightarrow 0$, the left-turning point is
 determined by the asymptotic limit
of
$U(y;{ m^2})$ as
$y\rightarrow -\infty$,
$U(y;{ m^2})
\rightarrow \epsilon  -({m^2/ 8}-{3/ 2} ){e^{y}}$. That is, 
$e^{y_L} \simeq {8\epsilon\over m^2-{12}}.$ As
$\epsilon \rightarrow 0^+$, for
${ m}$ large, it follows that the left-turning point moves to to minus
 infinity, $y_L\rightarrow -\infty$. More
explicitly, this is a consequence of the fact that $\tilde
L^2_T=\epsilon\rightarrow 0$. We  further note that, with  both $U_0(y)$ and
$U_1(y)$ vanish as
$0(e^y)$ as $y\rightarrow -\infty$, the ratio
$U_1(y)/U_0(y)$ remains finite; it is justified to expand the integrand in 
$m^{-2}$ and one obtains
\eqn\LeftExp{I_L(m) = \int_{-\infty}^{0} d y \sqrt
{-U_0(y)}+ 0({1\over m^2}) = \int_{0}^{1} d x \sqrt
{-V_0(x)} + 0({1\over m^2}).}

We turn next to  $I_R(m)$, which can be more conveniently written as
\eqn\RightI{I_R({ m}) = \int_{1}^{x_R} {d x\over x} 
 \sqrt {- x^2 V_0(x)} + 
\int_{1}^{x_R} {d x\over x}
{\sqrt {-x^2 V_0(x)}}
\Big\{\sqrt {1 +{1\over m^2}{\tilde V_1(x;m^2)\over 
{  V_0(x)}}} -1\Big\}}
The upper limit is determined by $\tilde V(x_R) =
 m^2\big( V_0(x_R) + {1\over m^2}
\tilde V_1(x_R)\big)=0$. Since
$V_0(x)
\rightarrow -{1/4x^3}$ and $\tilde V_1(x) 
\rightarrow {1\over x^2}$, it follows that 
$x_R(m) \simeq   {m^2/ 4}.$
The first integral  can be expressed as a difference:  
$\int_{1}^{\infty} d x \sqrt
{- V_0(x)}-\int_{x_R}^{\infty} d x
\sqrt {- V_0(x)} $, where, for ${1/ m}$ small, 
$\int_{x_R}^{\infty} d x
\sqrt {- V_0(x)} \simeq x_R^{-1/2}\simeq {2/ m}$. 
For the second integral, the main contribution now comes from region 
where $x =0( m^2)$. Let us
change variable from $x$ to $u\equiv {x/ x_R}\simeq 
{4x/ m^2}$, and, in the limit of large $m$,
this expression becomes
${1\over m} \int_{0}^{1} {d u\over u^{3/2}}\big( { \sqrt {1-u} }
-1\big ) + 0({1\over m^2})= ({2\over m})(1-{\pi\over
2})+ 0({1\over m^2})  . $ 
 Putting things together, we thus
have,
\eqn\RightExp{I_R(m)
=\int_{1}^{\infty} d x \sqrt {-V_0(x)} -{\pi\over m} + 
0({1\over m^2}).}
Combining this with that for $I_L$, we have, for tensor modes,
\eqn\FinalIT{I({m}) \simeq \int_{0}^{\infty} d x \sqrt {-V_0(x)} 
-{\pi\over m}+ 0({1\over m^2}) = {1\over
4} B({1\over 4}, {1\over 2}) - {\pi\over m} + 0({1\over m^2}).}
This leads to Eq.\LeadingCoeff\ for $\mu$, and $\delta_T=3$ for tensor modes.

Turning next to the vector modes. Again, separating $I(m)$ into two parts, 
Eq. \LR, an identical  analysis for
$I_R(m)$ can be carried out as was done for the tensor modes, leading to
 the expansion, Eq. \RightExp. However,
since $U(y)\rightarrow {1/ 4}$ as $y \rightarrow -\infty$, (corresponding to the
situation where $\tilde L_V=1/4$), Eq.
\LeftExp\ no longer holds for the vector modes. One finds, instead,
\eqn\LeftExpV{I_L(m) =\int_{0}^{1} d x \sqrt {-V_0(x)} 
-{\pi\over 2m} + 0({1\over m^2}).}
Combining this with that for $I_R$, we have, for vector modes,
\eqn\FinalIV{I({m}) \simeq \int_{0}^{\infty} d x 
\sqrt {-V_0(x)} -{3\pi\over 2m}+ 0({1\over m^2}) .}
This leads to Eq. \LeadingCoeff\ for $\mu$, and $\delta_V=4$ for vector modes.

We complete this analysis by treating the scalar modes. Since 
$\tilde L^2_S=\tilde L^2_T=0$, one finds that Eq.
\LeftExp\ holds for scalars modes also. As for $I_R(m)$, we need to work
with Eq. \RightI\ with $\tilde V(x;m^2)= {1/(4 x^2)} + V(x;m^2)$, where $V$ for
scalar
 modes is given by the rather involved
expression in 
\BPotential\ and \BPotentSS.
Let us consider the second integral first, where   the main 
contribution now comes from region where $x \sim
m^2$. Changing variable from $x$ to $u\equiv {x/m^2}$, and, 
in the limit of large $m$, this
expression becomes
$${1\over 2m} \int_{0}^{u_R} {d u\over u^{3/2}}\big( { \sqrt {1-3u^2 -9u^3} 
\over {1+3u}}
-1\big)+ 0({1\over m^2}), $$ 
where $u_R$ is the  positive root of the $1-3u^2 -9u^3=0$,  
$u_R\simeq .3915\>$.
 The first integral  can be expressed as a difference: 
 $\int_{1}^{\infty} d x \sqrt
{- V_0(x)}-\int_{x_R}^{\infty} d x
\sqrt {- V_0(x)} $, where, for ${1/ m}$ small, 
$\int_{x_R}^{\infty} d x
\sqrt {- V_0(x)} \simeq {1/(m \sqrt u_R)}.$
Combining these with that for $I_L$, we have
$$I({ m}) \simeq \int_{0}^{\infty} d x \sqrt {-V_0(x)} + {s_1\over m}
 + 0({1\over m^2}),$$
where 
$s_1\simeq -(2.018 )({\pi / 2})$, 
thus leading to 
$\delta_S=3.018\>$, Eq. \SDelta.

\listrefs\bye